\pdfoutput=1
\documentclass{article}
\usepackage{amsmath, amssymb}\usepackage{extarrows}
\usepackage{amsfonts}
\usepackage{appendix}
\usepackage{graphicx}
\usepackage{xcolor}
\usepackage{url}
\usepackage{bm}
\usepackage[all]{xy}
\usepackage{arydshln}
\usepackage{colortbl}
\usepackage{tabularx}
\usepackage{makecell}
\usepackage{textcomp}

\usepackage{multirow}
\usepackage{diagbox}
\usepackage{threeparttable}

\usepackage{cases}

\newtheorem{theorem}{Theorem}

\newtheorem{definition}{Definition}

\newcommand{\Z}{\ensuremath{\mathbb Z}}

\newcommand{\R}{\ensuremath{\mathbb R}}

\newcommand{\ls}[1]
{\dimen0=\fontdimen6\the\font\lineskip=#1\dimen0
	\advance\lineskip.5\fontdimen5\the\font
	\advance\lineskip-\dimen0
	\lineskiplimit=0.9\lineskip
	\baselineskip=\lineskip
	\advance\baselineskip\dimen0
	\normallineskip\lineskip\normallineskiplimit\lineskiplimit
	\normalbaselineskip\baselineskip
	\ignorespaces}

\begin{document}

\title{Boolean Functions of  Binary Type-II and Type-II/III Complementary Array Pair \\
}
\renewcommand{\thefootnote}{}
\footnotetext{The material in this paper was  presented in part at The 10th International Workshop on Signal Design and its Applications in Communications (IWSDA'2022).}
\author{Erzhong Xue$^{1}$, Zilong Wang$^{1}$, Jinjin Chai$^{2}$\\
	\small  $^1$State Key Laboratory of Integrated Service Networks, Xidian University \\[-0.8ex]
	\small Xi'an, 710071, China\\
			\small $^2$The Air Defence and Antimissile School, Air Force Engineering University \\
			\small Xi'an, 710100, China\\
	\small\tt 2524384374@qq.com, zlwang@xidian.edu.cn,  jj\_chai@163.com\\
}

\maketitle

\thispagestyle{plain} \setcounter{page}{1}

\ls{1.5}

\begin{abstract}

The sequence pairs of length $2^{m}$ projected from complementary array pairs of Type-II of size $\bm{2}^{(m)}$ and mixed Type-II/III    and of size $\bm{2}^{(m-1)}\times2$  are complementary sequence pairs Type-II and Type-III respectively.
An exhaustive search for binary Type-II and Type-III complementary sequence pairs of small lengths $2^{m}$ ($m=1,2,3,4$) shows that they are all projected from the aforementioned complementary array pairs,
whose algebraic normal forms satisfy specified expressions. 
It's natural to ask whether the conclusion holds for all $m$.
In this paper, we proved that  these  expressions of algebraic normal forms determine all the  binary complementary array pairs of Type-II of size $\bm{2}^{(m)}$ and mixed Type-II/III of size $\bm{2}^{(m-1)}\times2$ respectively.
\end{abstract}

{\bf Keywords: }
	Types I and II complementary pair, array, sequence, binary, Boolean function

\section{Introduction} \label{section:intro}
Golay complementary sequences were first introduced by Golay \cite{Golay1951Static} and had since found applications in diverse areas of digital communication such as channel measurement, synchronization, and power control for multi-carrier wireless transmission.

The binary Golay complementary sequences are known to exist for all lengths  $2^{a}10^{b}26^{c}$ (where $a$, $b$ and $c$ are natural numbers). For Golay complementary sequences of length $2^m$, their algebraic normal forms (ANFs) are given by Davis and Jedwab \cite{Davis1999Peak} in 1999.  These   sequences are called  \lq \lq standard\rq\rq Golay complementary sequences. It is still open now whether there are non-standard binary  Golay complementary sequences of length $2^m$.

One most powerful method to study Golay complementary sequences is Golay complementary arrays. The existence of a binary Golay complementary array of the given size was studied in \cite{Jedwab2007Golay}.
In addition, it has been shown in \cite{Fiedler2008Am} that all the known binary Golay complementary sequences of length $2^m$ can be obtained by a single binary Golay complementary array pair of dimension $m$ and size $2\times 2 \times \cdots \times 2$ (or abbreviated as $\bm{2}^{(m)}$). 

The  aforementioned Golay complementary sequence pairs (or array pairs) are referred to as Type-I complementary sequence pairs (or array pairs). Note that the Fourier spectrum of Type-I complementary polynomials are evaluated on the unit circle. By extending to evaluate them on the real axis and imaginary axis, respectively, 
Parker and Riera \cite{Parker2011Generalised} proposed Type-II and Type-III complementary arrays. For more research on Type-II and Type-III complementary arrays, readers can refer to the literature \cite{Parker2008Close,Bjorstad2009equivalence,Parker2009polynomial,Riera2010boolean,Chunlei2012Complementary}.

The existence and construction of binary Types-II and Types-III complementary sequence pairs  were further studied in \cite{Chunlei2012Complementary}. In particular, it was shown that the length of Type-II complementary sequence must be a power of $2$. 
Moreover, an exhaustive search in \cite{Chunlei2012Complementary} for binary Type-II and Type-III complementary sequence pairs for length $n=2^{m}$, $n = 2, 4, 8, 16 $ shows that Type-II complementary sequence pairs of these lengths must satisfy specified ANFs respectively. Similar to the Type-I case, an open question left in \cite{Chunlei2012Complementary} asked  whether all the  Type-II and Types-III complementary sequence pairs must satisfy these ANFs. 

Until now, all the known binary Types I and II complementary sequence pairs of length $2^m$ can be obtained by  binary Types I and II complementary array pairs of size $\bm{2}^{(m)}$ respectively,
all the known binary Type-III complementary sequence pairs of length $2^{m}$ can be obtained by  binary complementary array pairs, being of Type-III for the first variable, and Type-II for the other $m-1$ variables \cite{Chunlei2012Complementary}. We denote the complementary array pairs as mixed Type-II/III  of size $\bm{2}^{(m-1)}\times2$ in this paper. Thus, to study whether the Types I-III complementary sequence pairs must satisfy specified ANFs, one should first answer  whether the complementary array pairs of Types I-II of size $\bm{2}^{(m)}$ and mixed Type-II/III  of size  $\bm{2}^{(m-1)}\times2$ must satisfy these specified ANFs. The spectrum of Walsh-Hadamard transform and Nega-Walsh-Hadamard transform of Types I-III complementary array pairs has been determined in \cite{Chai2021DCCWalsh}.
In \cite{Type-I}, we proved that all $q$-ary Type-I complementary array pairs of size $\bm{2}^{(m)}$ must be standard.
In this paper, we proved that the ANFs of all the binary complementary array pairs of Type-II of size $\bm{2}^{(m)}$ and mixed Types-II/III  of size $\bm{2}^{(m-1)}\times2$ are determined by the form shown in \cite{Chunlei2012Complementary} respectively. 

The rest of the paper is organized as follows. In Section \ref{section:pre}, we introduce the Types I-III complementary sequences and arrays, and also introduce mixed Type-II/III arrays and projections from arrays to sequences. 
In Section \ref{Sec: main result}, our results for Type-II of size $\bm{2}^{(m)}$ and mixed Type-II/III array pairs of size $\bm{2}^{(m)}\times2$ are proposed.
We give the proof in 
Section~\ref{section:proof} and conclude the unresolved problem in Section~\ref{sec: Discussion}.

\section{Preliminary}\label{section:pre}
In this section, we introduce the basic notations and definitions of
Types I-III complementary sequences and arrays and mixed Type-II/III array.
We shall only study the binary case in  this paper.

\subsection{Type-I Complementary Sequence and Array}
A binary sequence $\boldsymbol{f}$ of length $L$ is defined as
\[\boldsymbol{f}=(f(0), f(1),\cdots, f(L-1)),\]
where each entry $f(t)$  belongs to  $\mathbb{Z}_2$ $(t\in \mathbb{Z}_{L})$.

\begin{definition}
	The {\em aperiodic auto-correlation} of binary sequence $\boldsymbol{f}$ at shift $\tau$ $(-L < \tau < L)$ is defined by 
	\[C_{\boldsymbol{f}}(\tau)=\sum_{t} (-1)^{{f}(t+\tau)-{f}(t)},\]
	where $(-1)^{{f}(t+\tau)-{f}(t)} := 0$ if ${f}(t+\tau)$ or ${f}(t)$ is not defined.
\end{definition}
\begin{definition}\label{def: GCP_C}
	A pair of sequences $\{\boldsymbol{f},\boldsymbol{g}\}$ is called a {\em Golay complementary pair} if
	$C_{{f}}(\tau)+C_{{g}}(\tau)=0$ for $\forall \tau \neq 0.$
	Each sequence in such a pair is called a {\em Golay complementary sequence} \cite{Golay1951Static}.
\end{definition}

The {\em generating function} of a binary sequence $f(t)$ is given by
\begin{equation}
	F({z})=\sum_{{t} \in \mathbb{Z}_L}(-1)^{f({t})}z^{t}.
\end{equation}
Straightforward manipulation shows that
\begin{equation}
	F({z})\cdot {F}({z}^{-1})=\sum_{{\tau}}C_{f}({\tau})z^{\tau}.
\end{equation}
Then we have that $f(t)$ and $g(t)$ are Golay complementary sequences if and only if their generating functions $F({z})$ and $G({z})$ satisfy
\begin{equation}\label{eq: GCP_F}
	F({z})\cdot
	{F}({z}^{-1})+G({z})\cdot {G}({z}^{-1})= 2{L}.
\end{equation}

An $m$-dimensional binary array of size $2 \times2\times\cdots\times2$ can be represented by a Boolean function
$$f(\boldsymbol{x})=f(x_1,x_2,\cdots,x_m):\{0,1\}^{m}\to \mathbb{Z}_2.$$
\begin{definition}\label{definition:d1}
	The {\em aperiodic auto-correlation} of an array $f(\boldsymbol{x})$ at shift $\boldsymbol{\tau} = (\tau_1,\tau_2,\cdots,\tau_m)$, $(-1\leq\tau_i\leq 1)$, is defined by
	\begin{equation*}\label{equation:eq1}
		C_f(\boldsymbol{\tau})=\sum_{\boldsymbol{x}\in \{0,1\}^m} (-1)^{f(\boldsymbol{x}+\boldsymbol{\tau})-f(\boldsymbol{x})},
	\end{equation*}
	where $(-1)^{f(\boldsymbol{x}+\boldsymbol{\tau})-f(\boldsymbol{x})}:=0$ if $f(\boldsymbol{x}+\boldsymbol{\tau})$ or $f(\boldsymbol{x})$ is not defined, and $\boldsymbol{x}+\boldsymbol{\tau}$ is the element-wise addition of integers.
\end{definition}

\begin{definition}\label{def: GAP_C}
	A pair of arrays $(f(\boldsymbol{x}),g(\boldsymbol{x}))$ is called a {\em Golay complementary array pair}  if
	\begin{equation*}\label{equation:eq2}
		C_{f}(\boldsymbol{\tau})+C_{g}(\boldsymbol{\tau})=0\ for\ \forall \boldsymbol{\tau} \neq \boldsymbol{0}.
	\end{equation*}
	Each array in such a pair is called a {\em Golay complementary array} \cite{Jedwab2007Golay}.
\end{definition}

The {\em generating function} of a binary array $f(\boldsymbol{x})$ is given by
\begin{equation}
	F(\boldsymbol{z})=\sum_{\boldsymbol{x} \in \{0,1\}^m}(-1)^{f(\boldsymbol{x})}z_1^{x_1}z_2^{x_2}\cdots z_m^{x_m},
\end{equation}
where $\boldsymbol{z}=(z_1,z_2,\cdots,z_m)$.

A Golay complementary array can be alternatively defined from the generating functions. Denote $\bm{z}^{-1}=(z_1^{-1},z_2^{-1},\cdots, z_{m}^{-1})$,
straightforward manipulation shows that
\begin{equation}
	F(\bm{z})\cdot {F}(\bm{z}^{-1})=\sum_{\bm{\tau}}C_{f}(\bm{\tau})z_1^{
		\tau_1}z_2^{\tau_2}\cdots z_{m}^{\tau_{m}}.
\end{equation}
From this it follows that $f(\bm{x})$ and $g(\bm{x})$ are Golay complementary arrays if and only if their generating functions $F(\bm{z})$ and $G(\bm{z})$ satisfy
\begin{equation}\label{eq: GAP_F}
	F(\bm{z})\cdot
	{F}(\bm{z}^{-1})+G(\bm{z})\cdot{G}(\bm{z}^{-1})= 2^{m+1}.
\end{equation}

\subsection{Types II and III Complementary Sequence and Array}

In 2008, Parker \cite{Parker2008Close,Parker2011Generalised} proposed Type-II and Type-III complementary sequences (or arrays).
Golay complementary sequence in Definition \ref{def: GCP_C} (or satisfying \eqref{eq: GCP_F}) is called Type-I complementary sequence, Golay complementary array in Definition \ref{def: GAP_C} (or satisfying (\ref{eq: GAP_F}) ) is called Type-I complementary array.
Just as each Type-I complementary polynomial is naturally evaluated on the unit circle to yield its Fourier
spectrum, Li et al. \cite{Chunlei2012Complementary} showed that it is natural to evaluate Type-II and Type-III complementary polynomials on
the real axis $\R$ and imaginary axis $\mathbb{I}$, respectively, to preserve the commutativity of conjugation in individual evaluations.

A pair of binary sequences  $\{\boldsymbol{f},\boldsymbol{g}\}$ of length $L$ is  called a Type-II and Type-III complementary sequence pair respectively if their generating functions  satisfy
\begin{equation}\label{eq: TypeII_seq}
	\begin{split}
		\text{Type-II sequence}:&
		\quad \frac{(F({z}))^{2}+(G({z}))^{2}}{1+z^2+z^4+\dots+z^{2(L-1)}}=2,
	\end{split}
\end{equation}
and
\begin{equation}\label{eq: TypeIII_seq}
	\begin{split}
		\text{Type-III sequence}:&\quad \frac{F({z})\cdot {F}(-{z})+G({z})\cdot {G}(-{z})}{1-z^2+z^4-\dots+(-1)^{L-1}z^{2(L-1)}}=2.
	\end{split}
\end{equation}

A  pair of binary arrays $(f(\boldsymbol{x}),g(\boldsymbol{x}))$ of size $2 \times2\times\cdots\times2$ is  called Type-II or Type-III complementary array pairs if their generating functions  satisfy
\begin{equation}\label{eq: TypeII_def}
	\text{Type-II array}:\quad
	(F(\bm{z}))^{2}+(G(\bm{z}))^{2}
	=2\prod_{k=1}^m(1+z_k^2),
\end{equation}
or
\begin{equation}\label{eq: TypeIII_def}
	\text{Type-III array}:\quad
	F(\bm{z})\cdot{F}(-\bm{z})+G(\bm{z})\cdot{G}(-\bm{z})
	=2\prod_{k=1}^m(1-z_k^2),
\end{equation}
respectively,
where 
$-\bm{z}=(-z_1,-z_2,\cdots,-z_{m})$.

\subsection{Mixed Type-II/III Complementary Array and Projections}
Suppose $f(t)$ is a sequence of length $L=2^{m}$,  $f(\bm{x})$ is an array of size $\bm{2}^{(m)}$, $F(z)$ and $F(\bm{z})$ are their corresponding generating functions.
The sequence $f(t)$ is called {\em projected} from the array $f(\bm{x})$ by permutation $\pi$ if 
\begin{equation}
	f(t)=f(\bm{x}),
\end{equation}
or equivalently,
\begin{equation}
F(z)=F(\bm{z}),
\end{equation}
where $t=\sum_{k=1}^{m}x_{k}\cdot2^{\pi(k)-1}$, and  $z_{k}=z^{2^{\pi(k)-1}}$,
$\pi$ is a permutation of $\{1,2,3,\dots,m\}$.
Notice that the pair of generating functions $(F(z),G(z))$ projected from that of the Type-I (resp. Type-II) complementary array pair $(F(\bm{z}),G(\bm{z}))$ in \eqref{eq: GAP_F} (resp. \eqref{eq: TypeII_def}) satisfy the condition of Type-I (resp. Type-II) complementary sequence pair in \eqref{eq: GCP_F} (resp. \eqref{eq: TypeII_seq}) for $L=2^{m}$.
It is straightforward that the sequence pair projected from Type-I (resp. Type-II) complementary array pair by permutation $\pi$ is a Type-I (resp. Type-II) complementary sequence pair.

However, the sequence pair projected from Type-III complementary array pair is not a Type-III complementary sequence pair.
As \cite{Chunlei2012Complementary} pointed out:

\noindent{\em All these Type-III sequence pairs of length $2^m$ are projections of m-variable $2\times2\times\cdots\times2$ bipolar array pairs, being of Type-III for the first variable, and Type-II for the other $m-1$ variables.}

\noindent
The aforementioned mixed type complementary array can be give the as  follows.

A  pair of binary arrays $(f(\bm{x},{x}_{0}),g(\bm{x},{x}_{0})$ is called a mixed Type-II/III complementary array pair of size $\bm{2}^{(m)}\times2$ if their generating functions  satisfy
\begin{equation}\label{eq: TypeII,III_def}
	F(\bm{z},{z}_{0})\cdot {F}(\bm{z},-{z}_{0})+G(\bm{z},{z}_{0})\cdot {G}(\bm{z},-{z}_{0})
	=2(1-z_{0}^2)\prod_{k=1}^{m}(1+z_k^2),
\end{equation}
where $z_{0}$ is called the Type-III indeterminate, and $x_{0}$ is its corresponding variable, $z_{k}$ ($1\leq{k}\leq{m}$) are called the Type-II indeterminates.

Notice that the pair of generating functions $(F(z),G(z))$ projected from that of the mixed Type-II/III complementary array pair $(F(\bm{z}),G(\bm{z}))$ in \eqref{eq: TypeII,III_def} satisfy the condition of Type-III complementary sequence pair in  \eqref{eq: TypeIII_seq} for $L=2^{m+1}$,
if we restrict $z_{0}=z$ and $z_{k}$ ($1\leq{k}\leq{m}$) to be $z^{2^{\pi(k)}}$, where $\pi$ is a permutation of $\{1,2,3,\dots,m\}$.
This means that the sequence pair projected from  mixed Type-II/III complementary array pair of size $\bm{2}^{(m)}\times2$ by a specific permutation is a Type-III complementary sequence pair.

\subsection{Some Known Results}
It has been proved that the length of Type-II complementary sequence must be a power of $2$ \cite{Chunlei2012Complementary}. In addition, 
an exhaustive search in \cite{Chunlei2012Complementary} for binary Type-II complementary sequence pairs $(f(t),g(t))$ of length $L=2^{m}$, $L = 2, 4, 8, 16 $, reveals that they are all projected from the Type-II complementary array pairs given by
\begin{numcases}{}
	f(\bm{x})=\sum_{1\leq i<j \leq m}x_ix_j+ \sum_{i=1}^{m}c_{i}x_{i}+c_{0},\label{eq: thm_f}\\
	g(\bm{x})=f(\bm x)+\sum_{i=1}^mx_i+c',\label{eq: thm_g}
\end{numcases}
where $c'\in\Z_{2}$, $c_{k}\in\Z_{2}$ $(0\leq k\leq m)$.
The projection can be expressed by
	\begin{equation}\label{eq: project_II}
		\left\{\begin{aligned}
			f(t)&=f(\bm{x}),\\
			g(t)&=g(\bm{x}),
		\end{aligned}\right.
	\end{equation}
	where $t=\sum_{k=1}^{m}2^{\pi(k)-1}\cdot{x}_{k}$,
	$\pi$ is a permutation of $\{1,2,3,\dots,m\}$.

An exhaustive search for the Type-III complementary sequence in \cite{Chunlei2012Complementary} reveals that, for length $L=2^{m+1}$ and
$L=2,4,8,16$, all binary Type-III complementary sequence pairs, $(f(t),g(t))$, are projected via \eqref{eq: project_III} from the Type-III complementary array pairs of size $\bm{2}^{(m)}\times2$ given by:
\begin{numcases}{}
f_{\text{II/III}}(\bm{x},{x}_{0})=\sum_{1\leq{i}<{j}\leq{m}}x_ix_j+ x_{0}\cdot\sum_{i=1}^{m}e_{i}x_{i}+\sum_{i=0}^{m}c_{i}x_{i}+c,\label{eq: f(x)_III}\\
g_{\text{II/III}}(\bm{x},{x}_{0})=f_{\text{II/III}}(\bm{x},{x}_{0})+\sum_{i=1}^{m}x_i+{e}_{0}x_{0}+c',\label{eq: g(x)_III}
\end{numcases}
where $c,c'\in\Z_{2}$, $e_{k},c_{k}\in\Z_{2}$ $(0\leq k\leq m)$.
The projection can be expressed by
	\begin{equation}\label{eq: project_III}
		\left\{\begin{aligned}
			f(t)&=f_{\text{II/III}}(\bm{x},{x}_{0}),\\
			g(t)&=g_{\text{II/III}}(\bm{x},{x}_{0}),
		\end{aligned}\right.
	\end{equation}
	where  $t=\sum_{k=1}^{m}2^{\pi(k)}\cdot{x}_{k}+x_{0}$, $\pi$ is a permutation of $\{1,2,3,\dots,m\}$.

\section{Main Results}\label{Sec: main result}

Based on the theoretical results of the lengths of the Type-II and Type-III complementary sequences and the exhaustive search for the Type-II and Type-III complementary sequence pairs of small lengths. 
Two open questions are proposed in \cite{Chunlei2012Complementary}:

{\em 1. Prove that all bipolar Type-II complementary sequence pairs are constructed from primitive pair ($ A = (1,1) $, $ B = (1,-1) $) by an $ m $-fold application of Construction $ G $, then a projection of the resulting $ m $-variate Type-II complementary array pair back to a sequence pair.}

{\em 2. Prove that all bipolar Type-III complementary sequence pairs of length $2^{m}$ can be constructed from primitive pair ($ A = (1, 1) $, $ B = (1,1) $) by an $ m $-fold application of Construction $ G $, then a projection of the resulting $ m $-variate Type-II/III complementary array pair back to a sequence pair.}

From the viewpoint of array and sequence, the two open problems can be reorganized into two parts. 
First, all binary Type-II (mixed Type-II/III) complementary arrays must be of form \eqref{eq: thm_f}, \eqref{eq: thm_g} (resp. \eqref{eq: f(x)_III}, \eqref{eq: g(x)_III}).
Second, all binary Type-II (resp. Type-III) complementary sequence pairs  of length $2^{m}$ are projected from these Type-II (resp. Type-II/III) complementary array pairs. In this paper, we will prove that the first part is true. 



\begin{theorem}\label{thm: Type IIA}
1. The array pair $(f(\bm{x}),g(\bm{x}))$ given in \eqref{eq: thm_f}, \eqref{eq: thm_g} form a Type-II complementary arrays of size $2\times2\times\cdots\times2$.

\noindent 2. Conversely, any Type-II complementary arrays $(f(\bm{x}),g(\bm{x}))$ of size $2\times2\times\cdots\times2$ must be of  form \eqref{eq: thm_f} and \eqref{eq: thm_g}.
\end{theorem}

\begin{theorem}\label{thm: Type III}
	1. The array pair $(f(\bm{x}),g(\bm{x}))$ given in \eqref{eq: f(x)_III} and \eqref{eq: g(x)_III} form a Type-II/III complementary arrays of size $\bm{2}^{(m)}\times2$.

	\noindent 2. Conversely, any Type-II/III complementary arrays $(f(\bm{x}),g(\bm{x}))$ of size $\bm{2}^{(m)}\times2$ must be of  form \eqref{eq: f(x)_III} and \eqref{eq: g(x)_III}.
\end{theorem}

\section{Proof of Our Results}\label{section:proof}

\subsection{Proof of Theorem \ref{thm: Type IIA}}
Define $F_{m+1}(\bm{z},z_{m+1})$ and $G_{m+1}(\bm{z},z_{m+1})$ as the generating functions of arrays $f_{m+1}(\bm{x},x_{m+1})$ and $ g_{m+1}(\bm{x},x_{m+1})$ of size $2^{(m+1)}$.
Denote $f_{m}^{t}(\bm{x})=f_{m+1}(\bm{x},t)$ ($t=0$ or $1$) (resp. $g_{m}^{t}(\bm{x})=g_{m+1}(\bm{x},t)$) by the array of dimension $m$ derived from $f_{m+1}(\bm{x},x_{m+1})$ (resp. $g_{m+1}(\bm{x},x_{m+1})$) by restricting $x_{m+1}$ to be $t$ ($t=0$ or $1$).
I.e.,
\begin{equation}
	f_{m+1}(\bm{x},x_{m+1})=f^{0}_{m}(\bm{x})(1-x_{m+1})+f^{1}_{m}(\bm{x})\cdot{x}_{m+1},
\end{equation}
\begin{equation}
	g_{m+1}(\bm{x},x_{m+1})=g^{0}_{m}(\bm{x})(1-x_{m+1})+g^{1}_{m}(\bm{x})\cdot{x}_{m+1}.
\end{equation}
Denote corresponding generating functions to be $F^{0}_{m}(\bm{z})$ (resp. $F^{1}_{m}(\bm{z})$).
It's easy to verify that
\begin{equation}\label{eq: Gm+1(z)}
	G_{m+1}(\bm{z},z_{m+1})=G^{0}_{m}(\bm{z})+G^{1}_{m}(\bm{z})\cdot{z}_{m+1},
\end{equation}
\begin{equation}\label{eq: Fm+1(z)}
	F_{m+1}(\bm{z},z_{m+1})=F^{0}_{m}(\bm{z})+F^{1}_{m}(\bm{z})\cdot{z}_{m+1}.
\end{equation}

We would like to give the proof by applying the mathematical induction.
It is know that Theorems \ref{thm: Type IIA} holds for $m=1,2,3$ and $4$.
Suppose Theorems \ref{thm: Type IIA} holds for $m$.
\\ \hspace*{\fill} \\
\noindent
\textbf{Step 1.}
	If $(f(\bm{x},x_{m+1}),g(\bm{x},x_{m+1}))$ are of form \eqref{eq: thm_f} and \eqref{eq: thm_g}, i.e.,
	\begin{equation}\label{eq: asser_f}
		f(\bm{x},x_{m+1})=\sum_{1\leq k<j \leq m+1}x_kx_j+ \sum_{k=1}^{m+1}c_{k}x_{k}+c_{0},
	\end{equation}
	and
	\begin{equation}\label{eq: asser_g}
		g(\bm{x},x_{m+1})=f(\bm{x},x_{m+1})+\sum_{k=1}^{m+1}x_k+c'.
	\end{equation}
	Then the restricted arrays are given by
\begin{numcases}{}
f_{m}^{0}(\bm{x})=\sum_{1\leq k<j \leq m}x_kx_j+ \sum_{k=1}^{m}c_{k}x_{k}+c_{0},\\
f_{m}^{1}(\bm{x})=f_{m}^{0}(\bm{x})+\sum_{k=1}^{m}x_{k}+c_{m+1},\\
g_{m}^{0}(\bm{x})=f_{m}^{1}(\bm{x})+c_{m+1}+c',\\
g_{m}^{1}(\bm{x})=f_{m}^{0}(\bm{x})+c_{m+1}+c'+1.
\end{numcases}
	So that $G^{0}_{m}(\bm{z})=\pm{F}^{1}_{m}(\bm{z})$ and $G^{1}_{m}(\bm{z})=\mp{F}^{0}_{m}(\bm{z})$.
	Since $(f_{m}^{0}(\bm{x}),g_{m}^{0}(\bm{x}))$ are of form \eqref{eq: thm_f} and \eqref{eq: thm_g}, they form Type-II complementary arrays.
	Their generating functions $(F_{m}^{0}(\bm{z}),G_{m}^{0}(\bm{z}))$ must satisfy \eqref{eq: TypeII_def}.
	Based on \eqref{eq: Gm+1(z)} and \eqref{eq: Fm+1(z)},
	\begin{equation}
		\begin{split}
			&\quad(F_{m+1}(\bm{z},z_{m+1}))^{2}+(G_{m+1}(\bm{z},z_{m+1}))^{2}\\
			&=((F_{m}^{0}(\bm{z}))^{2}+(G_{m}^{0}(\bm{z}))^{2})\cdot(1+z_{m+1}^2)\\
			&=2\prod_{k=1}^{m+1}(1+z_k^2),
		\end{split}
	\end{equation}
	which meets the definition of Type-II complementary arrays \eqref{eq: TypeII_def}. 
\\ \hspace*{\fill} \\
\noindent\textbf{Step 2.}
	If $F_{m+1}(\bm{z},z_{m+1})$ and $G_{m+1}(\bm{z},z_{m+1})$ form a Type-II complementary array pair, according to \eqref{eq: TypeII_def},
	\begin{equation}\label{eq: F^2+G^2}
		(F_{m+1}(\bm{z},z_{m+1}))^{2}+(G_{m+1}(\bm{z},z_{m+1}))^{2}=2\prod_{k=1}^{m+1}(1+z_k^2).
	\end{equation}
	On the other hand, from
	\eqref{eq: Gm+1(z)}$^{2}$ and \eqref{eq: Fm+1(z)}$^{2}$ we have
	\begin{equation}\label{eq: Fm+1^2}
		\begin{split}
			(F_{m+1}(\bm{z},z_{m+1}))^{2}&=(F^{0}_{m}(\bm{z}))^{2}+(F^{1}_{m}(\bm{z}))^{2}\cdot{z}_{m+1}^{2}\\
			&+2F^{0}_{m}(\bm{z})\cdot F^{1}_{m}(\bm{z})\cdot{z}_{m+1},
		\end{split}
	\end{equation}
	\begin{equation}\label{eq: Gm+1^2}
		\begin{split}
			(G_{m+1}(\bm{z},z_{m+1}))^{2}&=(G^{0}_{m}(\bm{z}))^{2}+(G^{1}_{m}(\bm{z}))^{2}\cdot{z}_{m+1}^{2}\\
			&+2G^{0}_{m}(\bm{z})\cdot G^{1}_{m}(\bm{z})\cdot{z}_{m+1}.
		\end{split}
	\end{equation}
	
	Expend the polynomial by the power of $z_{m+1}$, compare the coefficients of $1$, $z_{m+1}^{2}$ and $z_{m+1}$ respectively between \eqref{eq: Fm+1^2}+\eqref{eq: Gm+1^2} and \eqref{eq: F^2+G^2},
	we have
	\begin{equation}\label{eq: F0+G0}
		(F_{m}^{0}(\bm{z}))^{2}+(G_{m}^{0}(\bm{z}))^{2}=2\prod_{k=1}^{m}(1+z_k^2),
	\end{equation}
	\begin{equation}\label{eq: F1+G1}
		(F_{m}^{1}(\bm{z}))^{2}+(G_{m}^{1}(\bm{z}))^{2}=2\prod_{k=1}^{m}(1+z_k^2),
	\end{equation}
	\begin{equation}\label{eq: FF+GG=0}
		2(F_{m}^{0}(\bm{z})\cdot F_{m}^{1}(\bm{z})+G_{m}^{0}(\bm{z})\cdot{G}_{m}^{1}(\bm{z}))=0.
	\end{equation}
	Let
	\begin{equation}\label{eq: F=KG}
		F_{m}^{1}(\bm{z})= K(\bm{z})\cdot{G}_{m}^{0}(\bm{z}),
	\end{equation}
	where $K(\bm{z})$ belongs to the field of fractions of the polynomial ring.
	According to \eqref{eq: FF+GG=0}, we have
	\begin{equation}\label{eq: G=KF}
		G_{m}^{1}(\bm{z})=- K(\bm{z})\cdot{F}_{m}^{0}(\bm{z}).
	\end{equation} 
	Substituting $F_{m}^{1}(\bm{z})$ and $G_{m}^{1}(\bm{z})$ in \eqref{eq: F1+G1} by \eqref{eq: F=KG} and \eqref{eq: G=KF}, we have
	\begin{equation}\label{eq: K(F0+G0)}
		(K(\bm{z}))^{2}\cdot((F_{m}^{0}(\bm{z}))^{2}+(G_{m}^{0}(\bm{z}))^{2})=2\prod_{k=1}^{m}(1+z_k^2).
	\end{equation}
	Compare \eqref{eq: K(F0+G0)} with \eqref{eq: F0+G0}, we get
	\begin{equation}
		K(\bm{z})=\pm1.
	\end{equation}
	If $K(\bm{z})=1$, based on \eqref{eq: F=KG} and \eqref{eq: G=KF}, it's easy to know
	\begin{equation}\label{eq: f=kg}
		f_{m}^{1}(\bm{x})={g}_{m}^{0}(\bm{x}),~
		g_{m}^{1}(\bm{x})={f}_{m}^{0}(\bm{x})+1,
	\end{equation}
	According to \eqref{eq: F0+G0},
	$F_{m}^{0}(\bm{z})$ and $G_{m}^{0}(\bm{z})$ form a Type-II complementary array pair. Since Theorem \ref{thm: Type IIA} holds for $m$,
	let
	\begin{equation}
		{f}_{m}^{0}(\bm{x})=\sum_{1\leq k<j \leq m}x_kx_j+ \sum_{k=1}^{m}c_{k}x_{k}+c_{0},
	\end{equation}
	\begin{equation}
		{g}_{m}^{0}(\bm{x})={f}_{m}^{0}(\bm{x})+\sum_{k=1}^mx_k+c_{m+1}.
	\end{equation}
	where $c_{k}\in\Z_{2}$ $(0\leq k\leq m+1)$.
	Then
	\begin{equation}
		\begin{split}
			f_{m+1}&(\bm{x},x_{m+1})=f^{0}_{m}(\bm{x})(1-x_{m+1})+f^{1}_{m}(\bm{x})\cdot{x}_{m+1}\\
			&=f^{0}_{m}(\bm{x})+x_{m+1}\cdot\left(\sum_{k=1}^{m}x_k+c_{m+1}\right)\\
			&=\sum_{1\leq k<j \leq m+1}x_kx_j+ \sum_{k=1}^{m+1}c_{k}x_{k}+c_{0},
		\end{split}
	\end{equation}
	
	\begin{equation}
		\begin{split}
			g_{m+1}&(\bm{x},x_{m+1})=g^{0}_{m}(\bm{x})(1-x_{m+1})+g^{1}_{m}(\bm{x})\cdot{x}_{m+1}\\
			&={f}_{m}^{0}(\bm{x})+\left(\sum_{k=1}^mx_k+c_{m+1}\right)(1+x_{m+1})+x_{m+1}\\
			&=\sum_{1\leq k<j \leq m+1}x_kx_j+ \sum_{k=1}^{m+1}c_{k}x_{k}+\sum_{k=1}^{m+1}x_k+c_{0}+c_{m+1},
		\end{split}
	\end{equation}
	which are obviously of form \eqref{eq: thm_f} and \eqref{eq: thm_g}.
	If $k=-1$, we can get the similar result.

Combine Steps 1 and 2, Theorem \ref{thm: Type IIA} holds for $m+1$.
By applying the mathematical induction, Theorem \ref{thm: Type IIA} holds for all $m\ge1$.
	
\subsection{Proof of Theorem \ref{thm: Type III}}

Define $F_{\text{II/III}}(\bm{z},{z}_{0})$ and $G_{\text{II/III}}(\bm{z},{z}_{0})$ as the generating functions of 
arrays $f_{\text{II/III}}(\bm{x},x_{0})$ and $ g_{\text{II/III}}(\bm{x},x_{0})$, 
where $z_{0}$ is the  Type-III indeterminate, $z_{k}$ ($1\leq{k}\leq{m}$) are the Type-II indeterminates.

Denote $f_{m}^{t}(\bm{x})=f_{\text{II/III}}(\bm{x},t)$ ($t=0$ or $1$) (resp. $g_{m}^{t}(\bm{x})=g_{\text{II/III}}(\bm{x},t)$) by the 
array of dimension $m$ derived from $f_{\text{II/III}}(\bm{x},{x}_{0})$ (resp. $g_{\text{II/III}}(\bm{x},{x}_{0})$) by restricting ${x}_{0}$ to be $t$ ($t=0$ or $1$).
Thus,
\begin{equation}
	f_{\text{II/III}}(\bm{x},{x}_{0})=f^{0}_{m}(\bm{x})(1-{x}_{0})+f^{1}_{m}(\bm{x})\cdot{x}_{0},
\end{equation}
\begin{equation}
	g_{\text{II/III}}(\bm{x},{x}_{0})=g^{0}_{m}(\bm{x})(1-{x}_{0})+g^{1}_{m}(\bm{x})\cdot{x}_{0}.
\end{equation}
Denote corresponding generating functions to be $F^{t}_{m}(\bm{z})$ ($t=0$ or $1$) (resp. $G^{t}_{m}(\bm{z})$).
It's easy to verify that
\begin{equation}\label{eq: F23(z)}
	{F}_{\text{II/III}}(\bm{z},z_{0})=F^{0}_{m}(\bm{z})+F^{1}_{m}(\bm{z})\cdot{z}_{0},
\end{equation}
\begin{equation}\label{eq: G23(z)}
	{G}_{\text{II/III}}(\bm{z},z_{0})=G^{0}_{m}(\bm{z})+G^{1}_{m}(\bm{z})\cdot{z}_{0}.
\end{equation}
\\ \hspace*{\fill} \\
\noindent
\textbf{Step 1.}
If $(f_{\text{II/III}}(\bm{x},{x}_{0}),g_{\text{II/III}}(\bm{x},{x}_{0}))$ are given by \eqref{eq: f(x)_III} and \eqref{eq: g(x)_III} respectively. Then the restricted arrays are given by
\begin{numcases}{}
f_{m}^{0}(\bm{x})=\sum_{1\leq{i}<{j}\leq{m}}x_ix_j+ \sum_{i=1}^{m}c_{i}x_{i}+c,\\
{f}_{m}^{1}(\bm{x})=f_{m}^{0}(\bm{x})+ \sum_{i=1}^{m}e_{i}x_{i}+c_{0},\\
g_{m}^{0}(\bm{x})=f_{m}^{0}(\bm{x})+\sum_{i=1}^{m}x_i+c',\\
g_{m}^{1}(\bm{x})=f_{m}^{1}(\bm{x})+\sum_{i=1}^{m}x_i+{e}_{0}+c'.
\end{numcases}

Since $(f_{m}^{0}(\bm{x}),g_{m}^{0}(\bm{x}))$ are of form \eqref{eq: thm_f} and \eqref{eq: thm_g}, they form Type-II complementary arrays.
	Their generating functions $(F_{m}^{0}(\bm{z}),G_{m}^{0}(\bm{z}))$ must satisfy \eqref{eq: TypeII_def}, i.e.,
\begin{equation}
	(F_{m}^{0}(\bm{z}))^{2}+(G_{m}^{0}(\bm{z}))^{2}
	=2\prod_{k=1}^m(1+z_k^2),
\end{equation}
Similarly,
\begin{equation}
	(F_{m}^{1}(\bm{z}))^{2}+(G_{m}^{1}(\bm{z}))^{2}
	=2\prod_{k=1}^m(1+z_k^2),
\end{equation}

	Based on \eqref{eq: G23(z)} and \eqref{eq: F23(z)},
	\begin{equation}
		\begin{split}
			&\quad
			{F}_{\text{II/III}}(\bm{z},z_{0})\cdot{F}_{\text{II/III}}(\bm{z},-z_{0})
			+{G}_{\text{II/III}}(\bm{z},z_{0})\cdot{G}_{\text{II/III}}(\bm{z},-z_{0})\\
			&=((F_{m}^{0}(\bm{z}))^{2}+(G_{m}^{0}(\bm{z}))^{2})+z_{0}^2\cdot
			((F_{m}^{1}(\bm{z}))^{2}+(G_{m}^{1}(\bm{z}))^{2})\\
			&=2(1-z_0^2)\prod_{k=1}^{m}(1+z_k^2),
		\end{split}
	\end{equation}
	which meets the definition of Type-II/III complementary arrays \eqref{eq: TypeII_def} of size $\bm{2}^{(m)}\times2$. 
\\ \hspace*{\fill} \\
\noindent
\textbf{Step 2.}
	If ${F}_{\text{II/III}}(\bm{z},z_{0})$ and ${G}_{\text{II/III}}(\bm{z},z_{0})$ form a Type-II/III complementary array pair of size $\bm{2}^{(m)}\times2$. According to \eqref{eq: TypeII_def},
	\begin{equation}\label{eq: FF(-z)+GG(-z)}
	{F}_{\text{II/III}}(\bm{z},z_{0})\cdot{F}_{\text{II/III}}(\bm{z},-z_{0})
	+{G}_{\text{II/III}}(\bm{z},z_{0})\cdot{G}_{\text{II/III}}(\bm{z},-z_{0})=2(1-z_{0}^2)\cdot\prod_{k=1}^{m}(1+z_k^2).
	\end{equation}
	On the other hand, from
	\eqref{eq: G23(z)} and \eqref{eq: F23(z)} times their individual conjugates, we have
	\begin{equation}\label{eq: F23^2}
		\begin{split}
		{F}_{\text{II/III}}(\bm{z},z_{0})\cdot{F}_{\text{II/III}}(\bm{z},-z_{0})&=(F^{0}_{m}(\bm{z}))^{2}-(F^{1}_{m}(\bm{z}))^{2}\cdot{z}_{0}^{2},
		\end{split}
	\end{equation}
	\begin{equation}\label{eq: G23^2}
		\begin{split}
		{G}_{\text{II/III}}(\bm{z},z_{0})\cdot{G}_{\text{II/III}}(\bm{z},-z_{0})&=(G^{0}_{m}(\bm{z}))^{2}-(G^{1}_{m}(\bm{z}))^{2}\cdot{z}_{0}^{2}.
		\end{split}
	\end{equation}
	
	Expend the polynomial by the power of $z_{0}$, compare the coefficients of $1$, $z_{0}^{2}$ and $z_{0}$ respectively between \eqref{eq: F23^2}+\eqref{eq: G23^2} and \eqref{eq: FF(-z)+GG(-z)},
	we have
	\begin{equation}\label{eq: F0^2+G0^2}
		(F_{m}^{0}(\bm{z}))^{2}+(G_{m}^{0}(\bm{z}))^{2}=2\prod_{k=1}^{m}(1+z_k^2),
	\end{equation}
	\begin{equation}\label{eq: F1^2+G1^2}
		(F_{m}^{1}(\bm{z}))^{2}+(G_{m}^{1}(\bm{z}))^{2}=2\prod_{k=1}^{m}(1+z_k^2),
	\end{equation}

	According to \eqref{eq: TypeII_def},
	$(F_{m}^{0}(\bm{z}), G_{m}^{0}(\bm{z}))$ and $(F_{m}^{1}(\bm{z}), G_{m}^{1}(\bm{z}))$form Type-II complementary array pairs. According to Theorem \ref{thm: Type IIA},
	let
	\begin{numcases}{}
	{f}_{m}^{0}(\bm{x})=\sum_{1\leq i<j \leq m}x_ix_j+ \sum_{i=1}^{m}c_{i}x_{i}+c_{0},\\
	{g}_{m}^{0}(\bm{x})=f(\bm x)+\sum_{i=1}^mx_i+c',
	\end{numcases}
\begin{numcases}{}
	{f}_{m}^{1}(\bm{x})=\sum_{1\leq i<j \leq m}x_ix_j+ \sum_{i=1}^{m}e_{i}x_{i}+e_{0},\\
	{g}_{m}^{1}(\bm{x})=f(\bm x)+\sum_{i=1}^mx_i+e',
\end{numcases}
	where $c',e'\in\Z_{2}$, $c_{i},e_{i}\in\Z_{2}$ $(0\leq i\leq m)$.
	Then
\begin{numcases}{}
{f}_{\text{II/III}}(\bm{x},{x}_{0})=f^{0}_{m}(\bm{x})(1-{x}_{0})+f^{1}_{m}(\bm{x})\cdot{x}_{0}\notag\\
\qquad\qquad\quad\;\,\,=\sum_{1\leq{i}<{j}\leq{m}}x_ix_j+ x_{0}\cdot\sum_{i=1}^{m}(e_{i}-c_{i})x_{i}+\sum_{i=1}^{m}c_{i}x_{i}+(e_{0}-c_{0})x_{0}+c_{0},\\
g_{\text{II/III}}(\bm{x},{x}_{0})=g^{0}_{m}(\bm{x})(1-{x}_{0})+g^{1}_{m}(\bm{x})\cdot{x}_{0}
=f_{\text{II/III}}(\bm{x},{x}_{0})+\sum_{i=1}^mx_i+(e'-c')x_{0}+c',
\end{numcases}
	which are obviously of form \eqref{eq: f(x)_III} and \eqref{eq: g(x)_III}.

\section{Conclusion}\label{sec: Discussion}

In this paper, we proved that the algebraic normal forms of binary Type-II of size $\bm{2}^{(m)}$ and mixed Type-II/III complementary array pairs of size $\bm{2}^{(m-1)}\times2$ must satisfy specified expressions.
The unresolved problem is that whether the Type-II and Type-III complementary sequence pairs of length $2^{m}$ must be projected from these array pairs. And we left it as an open problem.


\begin{thebibliography}{00}


\bibitem{Bjorstad2009equivalence}
T. E. Bj{\o}rstad, M. G. Parker, and S. Center,
\newblock \lq\lq Equivalence Between Certain
Complementary Pairs of types I and III,\rq\rq 
In {\em Enhancing Cryptographic
	Primitives with Techniques from Error Correcting Codes},
volume 23 of {\em NATO Science for Peace and Security Series - D: Information and Communication Security}, no. 8, pages 203--221, IOS Press, 2009.


\bibitem{Chai2021DCCWalsh}
J. Chai, Z. Wang, and E. Xue,
\newblock \lq\lq Walsh Spectrum and Nega Spectrum of Complementary Arrays,\rq\rq \
{\em Designs Codes and Cryptography},
vol. 89, pp. 2663--2677, 2021.	


\bibitem{Davis1999Peak}
J. A. Davis and J. Jedwab,
\newblock \lq\lq Peak-to-mean power control in OFDM, Golay complementary sequences, and Reed-Muller codes,\rq\rq \
{\em IEEE Trans. Inf. Theory},
vol. 45, no. 7, pp. 2397--2417, 1999.	



\bibitem{Fiedler2008Am}
F. Fiedler, J. Jedwab, and M. G. Parker,
\newblock \lq\lq A multi-dimensional approach to the construction and enumeration of Golay complementary sequences,\rq\rq 
{\em Journal of Combinatorial Theory, Series A},
vol. 115, no. 5, pp. 753--776, 2008.	



\bibitem{Golay1951Static}
M. J. Golay,
\newblock \lq\lq static multislit spectrometry and its application to the panoramic display of infrared spectra,\rq\rq 
{\em Journal of the Optical Society of America},
vol. 47, no. 7, pp. 468--472, 1951.	


\bibitem{Jedwab2007Golay}
J. Jedwab and M. G. Parker,
\newblock \lq\lq Golay complementary array pairs,\rq\rq \
{\em Designs Codes and Cryptography},
vol. 44, no. 7, pp. 209--216, 2007.	



\bibitem{Chunlei2012Complementary}
C. Li, N. Li, and M. G. Parker,
\newblock \lq\lq Complementary sequence pairs of types I and III,\rq\rq \
{\em IEICE Transactions on Fundamentals of Electronics,
	Communications and Computer Sciences},
vol. E95-A, no. 11, pp. 1819--1826, 2012.	



\bibitem{Parker2008Close}
M. G. Parker,
\newblock \lq\lq Close encounters with Boolean functions of three different kinds,\rq\rq \
In {\em Coding Theory and Applications, Second International Castle Meeting, ICMCTA 2008, Castillo de la Mota, Medina del Campo, Spain, September 15-19, 2008. Proceedings},
pp. 137--153, 2008.	


\bibitem{Parker2009polynomial}
M. G. Parker,
\newblock \lq\lq Polynomial residue systems via unitary transforms,\rq\rq \
In {\em Invited talk in Post-Proceedings of Contact Forum Coding Theory and Cryptography III, The Royal Flemish Acadamy of Belgium for Science and the Arts, Brussels, Belgium},
2009.	


\bibitem{Parker2011Generalised}
M. G. Parker and C. Riera,
In \newblock \lq\lq Generalised complementary arrays,\rq\rq \
{\em Cryptography and Coding - 13th IMA International Conference, IMACC 2011, Oxford, UK, December 12-15, 2011. Proceedings},
Lecture Notes in Computer Science, pp. 41--60, Springer, 2011.


\bibitem{Riera2010boolean}
C. Riera and M. G. Parker,
\newblock \lq\lq Boolean functions whose restrictions are highly nonlinear,\rq\rq \
In {\em 2010 IEEE Information Theory Workshop, ITW 2010, Dublin, Ireland, August 30 - September 3, 2010},
no. 8, pp. 1--5, IEEE, 2010.	

\bibitem{Type-I}
E. Xue, and Z. Wang,
\newblock \lq\lq The $q$-ary Golay complementary arrays of size $\bm{2}^{(m)}$ are standard,\rq\rq
[Online]. Available: https://arxiv.org/abs/2207.04374
\end{thebibliography}

\end{document}